\documentclass{svjour3}
\smartqed
\usepackage{graphicx}
\usepackage{exscale}
\usepackage[intlimits]{amsmath}
\usepackage{amsfonts}
\usepackage{amssymb,amscd}
\usepackage{epsfig}
\usepackage{color}

\newcommand*{\no}{\noindent}
\newcommand*{\bea}{\begin{eqnarray}}
\newcommand*{\eea}{\end{eqnarray}}
\newcommand*{\be}{\begin{equation}}
\newcommand*{\ee}{\end{equation}}
\newcommand*{\pd}{\partial}

\newcommand*{\pref}[1]{(\ref{#1})}

\newcommand*{\nn}{\nonumber}

\newcommand*{\bma}{\begin{matrix}}
\newcommand*{\ema}{\end{matrix}}
\newcommand*{\bpm}{\left(\bma}
\newcommand*{\epm}{\ema\right)}

\begin{document}

\title{Truncating first-order Dyson-Schwinger equations in Coulomb-Gauge Yang-Mills theory}

\author{Reinhard Alkofer \and Axel Maas \and Daniel Zwanziger}

\institute{Reinhard Alkofer \and Axel Maas \at Institute for Physics, Karl-Franzens University Graz, Universit\"atsplatz 5, A-8010 Graz, Austria \and Daniel Zwanziger \at New York University, New York, NY 10003, USA}

\date{Received: date / Accepted: date}

\maketitle

\begin{abstract}
The non-perturbative domain of QCD contains confinement, chiral symmetry breaking, and the bound state spectrum. For the calculation of the latter, the Coulomb gauge is particularly well-suited. Access to these non-perturbative properties should be possible by means of the Green's functions. However, Coulomb gauge is also very involved, and thus hard to tackle.  We introduce a novel BRST-type operator $r$, and show that the left-hand side of Gauss' law is $r$-exact. 

We investigate a possible truncation scheme of the Dyson-Schwinger equations in first-order formalism for the propagators based on an instantaneous approximation. We demonstrate that this is insufficient to obtain solutions with the expected property of a linear-rising Coulomb potential. We also show systematically that a class of possible vertex dressings does not change this result.

\end{abstract}

\section{Introduction}

QCD in Coulomb gauge has a long history as a gauge being particularly useful for the calculation of the bound-state spectrum, one of the most intricate properties of QCD. Furthermore, a well-developed confinement scenario exists in the Coulomb gauge that was developed by Gribov and elaborated by one of us \cite{Gribov:1977wm,Zwanziger:1998ez}. Recently, it has also been shown that Coulomb gauge furnishes an upper limit for the physical string-tension and static quark-anti-quark potential already at the level of the two-point functions~\cite{Zwanziger:2002sh,Zwanziger:2003de,Cucchieri:2000hv}. Coulomb gauge can also be understood as a gauge which in a certain sense is physical: Unlike e.g.\ in covariant gauges, perturbatively unphysical gauge-fixing degrees of freedom do not propagate. Finally, its structure makes it especially suited for calculations at finite temperature. The results obtained in this gauge can also provide input to model calculations (see e.\ g.\ \cite{Szczepaniak:2001rg}) in the form of potentials, and is by this relevant to the work (e.\ g.\ \cite{Melde:2008yr}) of the jubilee to whom this volume and this work is dedicated.

These are particular advantages of the Coulomb gauge compared to, e.g., covariant gauges. However, these properties do come at the price of loosing explicit covariance, and therefore introducing a much more complicated structure in the mathematical description. As a consequence, the question of renormalizability \cite{Zwanziger:1998ez,Baulieu:1998kx,Watson:2008fb,watsondses} of this gauge is a very hard problem. Furthermore, the lack of covariance makes explicit calculations cumbersome even in perturbation theory \cite{Watson:2008fb,watsondses}.

Therefore, the determination of the Green's functions in the perturbative and non-perturbative domain is far less developed than in covariant gauges. Nonetheless, we will present an attempt at such a calculation in the far infrared regions by means of a method which has proven to be rather useful in Landau gauge \cite{Alkofer:2000wg,Fischer:2006ub,Alkofer:2004it,Fischer:2006vf}, an asymptotic analysis in terms of critical Green's functions. We attempt no full solution of the infinite tower of Green's functions, as is possible in Landau gauge \cite{Alkofer:2004it,Fischer:2006vf}, and will also restrict to the case of Yang-Mills theory, thus neglecting quarks. For this purpose, the first-order formulation will be used. A possible translation to the more widespread used second-order formalism can, in principle, be made using the translation prescription discussed in \cite{VillalbaChavez:2008dv}.

It is a remarkable fact that such calculations in Coulomb gauge in two dimensions can be done analytically \cite{Reinhardt:2008ij}. The results of these calculations confirm the scenario, which will be used throughout this work. This result is, of course, due to the strong constraints the Coulomb gauge imposes specifically in two dimensions. In higher dimensions, the situation is much more complex, and many attempts to obtain the Green's functions \cite{Alkofer:1988tc}, in particular using variational principles \cite{Feuchter:2004mk} and lattice calculations \cite{Cucchieri:2000kw,Cucchieri:2007uj,Nakagawa:2009zf,Burgio:2008jr}, have been performed.

We proceed as follows. General aspects of the Coulomb gauge and the first-order formalism are discussed in section (\ref{spsa}), and the corresponding DSEs in section (\ref{sdses}).  In section (2) we also introduce a novel on-shell BRST-type operator $r$ and show that the left-hand side of Gauss' law is $r$-exact.  The truncation employed here, described in section (\ref{strunc}), provide a consistent result at the level of power-counting. Unfortunately, if a linear rising Coulomb potential is required, this solution ceases to exist, as is discussed in detail in section (\ref{scrit}). This persists even when including consistent vertex modifications, as is discussed in section (\ref{svbt}). The results provide an example where a solution exists by power counting, but is not a solution of the system. Of course, this may be an artifact of truncation. The results of the investigation are then summarized in section (\ref{ssum}).

\section{General aspects of Coulomb gauge in the first order formalism}\label{spsa}

\subsection{Phase-space action}

The local Euclidean action in Coulomb gauge in phase-space or first-order formalism in $s$ spatial dimensions is
\be
S  =  \int d^{s+1}x  \Big[ i \pi_i (\pd_0 A_i - D_i A_0) + { 1 \over 2} (\pi^2 + B^2)   + \pd_i \bar{c} D_i c + i\pd_i b A_i \Big],\nn
\ee
\no where
\be
B_i^a = B_i^a(A)= \epsilon_{ijk}(\pd_j A_k + {1\over2}gA_j  \times A_k )\nn
\ee
\no is the color-magnetic field, and
\be
(A \times B)^a \equiv f^{abc}A^bB^c.\nn
\ee
\no Here $c$ and $\bar{c}$ are the ghost pair, and $D_i = D_i(A)$ is the gauge covariant derivative, $D_i c = \pd_i c + gA_i \times c$. All fields are understood to carry a color index in the adjoint representation that is contracted.  The field $b$ is a Lagrange multiplier that enforces the Coulomb gauge condition $\pd_i A_i = 0$. Here $\pi_i$ is an auxiliary field that represents an independent color-electric field.

The Euclidean phase-space measure is
\be
\int dA_i\, dA_0\, d\pi\, dc\, d \bar{c}\, db \exp(- S).\nn
\ee
\no If one integrates out $\pi_i$, one gets the second-order or Lagrangian Euclidean action in Coulomb gauge.

As in the Landau gauge \cite{Fischer:2008uz}, it will be assumed henceforth that a non-per\-tur\-ba\-tive realization of the Coulomb gauge exists which furnishes a globally well-defined and unbroken BRST symmetry\footnote{The problem of the existence of a globally well-defined BRST charge in this gauge is not yet solved, and similar considerations as in Landau gauge apply \cite{Fischer:2008uz,vonSmekal:2007}. However, a construction as proposed in Landau gauge \cite{vonSmekal:2007,vonSmekal:2008ws} may also be possible in Coulomb gauge. Furthermore, the existence of a conserved and vanishing global color charge \cite{Reinhardt:2008pr} may also be helpful in this respect.}. If this is the case, an analogue to the classical Gauss' law can be formulated using the BRST symmetry, as explained below.  As a consequence, certain relations between the renormalization constants of Coulomb gauge have to hold, in particular $Z_{A_0} = Z_c$, relating the ghost renormalization with that of the temporal gluon propagator \cite{Zwanziger:1998ez}.

\subsection{On-shell action and on-shell BRST operator}

We integrate out the auxiliary $b$-field, so that the gauge condition $\pd_i A_i = 0$ is satisfied identically (on-shell). Writing $A_i = A^\mathrm{T}_i$,  the on-shell action 
\be
S_{\rm os} =  \int d^{s+1}x  \Big[ i \pi_i (\pd_0 A^\mathrm{T}_i - D_i A_0) + { 1 \over 2} (\pi^2 + B^2) + \pd_i \bar{c} D_i c  \Big],\nn
\ee
\no is obtained, where $B = B(A^\mathrm{T})$, and $D = D(A^\mathrm{T})$.

Since the gauge condition is satisfied identically, it is appropriate to restrict the integration over $A^\mathrm{T}$ to the fundamental modular region $\Lambda$,  a region free of Gribov copies.  The other variables are unrestricted, and the path integral is given by
\be
\label{fmr}
\int_\Lambda dA^\mathrm{T}\, d\pi\, dA_0\, dc\, d\bar{c} \exp(-S_{\rm os}).
\ee
\no In the phase-space action, $A_0$ appears as a Lagrange multiplier, and if one integrates out $A_0$, one obtains $\delta(D_i \pi_i)$, which imposes Gauss' law\footnote{In the presence of quarks one would have instead $\delta(D_i \pi_i - \rho_{\rm qu})$.} $D_i \pi_i = 0$. Note that the integrand vanishes on the boundary of the Gribov region, after integrating out the $\pi$-field, because the Faddeev-Popov determinant vanishes there, and hence the Dyson-Schwinger equations (DSEs) are unchanged by a cut-off there \cite{Zwanziger:2001kw}.

For purposes of deriving the DSEs, it is preferable not to integrate out $A_0$ because that would produce a non-polynomial action.  Instead we exhibit an on-shell BRST-type operator which encodes Gauss' law, and provides a definition of physical observables\footnote{We note that this operator is affected by the same lack of knowledge concerning its global definition as in case of the BRST symmetry. Once more, we assume that it is globally well-defined.}. The on-shell BRST-type operator $r$ is defined by
\bea
r \bar{c}  &=  i D_i \pi_i;\quad\quad\quad\quad\quad r A_i^\mathrm{T}  &=  r \pi_i = 0;\label{roncbar}\\
rA_0  &=  \pd_i D_i c;\quad\quad\quad\quad\quad rc &= 0.\label{ronazero}
\eea
An operator similar to $r$ was introduced in eq.~(143) of \cite{Baulieu:1998kx}. These equations imply that $r$ is nilpotent, $r^2 = 0$. Moreover  it is a symmetry of $S_{\rm os}$,
\be
rS_{\rm os} = 0,\nn
\ee
as one sees by writing the on-shell action in the form
\be	
S_{\rm os}  =   \int d^4x \Big[   i \pi_i  \pd_0 A_i^\mathrm{T}  
+ { 1 \over 2}( \pi^2 + B^2 )
 + r( \bar{c} A_0 )  \Big].	\nn
\ee
The $r$-symmetry is compatible with the restriction to the fundamental modular region~$\Lambda$ of the integral~(\ref{fmr}) over $A^\mathrm{T}$ because $rA^\mathrm{T} = 0$. 

We identify physical observables with the cohomology of~$r$. With this definition of observables, one sees from (\ref{roncbar})  and (\ref{ronazero}) that the variables $A_i^\mathrm{T}$ and $\pi_i$ are physical, while $(A_0, c, \bar{c}, D_i\pi_i)$ form an unphysical quartet.  Any $r$-exact quantity $rX$ has vanishing expectation value $\langle rX \rangle = 0$, and is physically equivalent to 0, $rX \sim 0$.  Gauss' law is encoded in the $r$-symmetry because, according to~(\ref{roncbar}), $D_i\pi_i$ is an $r$-exact quantity, and is thus equivalent to~0, $D_i\pi_i \sim 0$.  The $r$-symmetry is not as powerful as the BRST symmetry because it does not determine the part of the action that is not $r$-exact.  However it does imply the cancellation of the instantaneous ghost and scalar loops which contain the notorious energy divergences of the Coulomb gauge. 

In the BRST formalism $A^{\rm T}_i$ and $\pi_i$ are not physical. However here they can be because the gauge is fixed on-shell. To make contact with the usual Coulomb gauge, in which only the transverse degrees of freedom are considered physical, we separate the $\pi$-field into transverse and longitudinal parts, $\pi_i = \pi_i^{\rm T} - \pd_i \phi$, where $\phi$ is the color-Coulomb potential, and $\pd_i \pi_i^{\rm T} = 0$. These fields satisfy
\bea
&r A^\mathrm{T}_i =  r \pi^\mathrm{T}_i = 0&\label{rontransva}\\
&r (M^{-1}\bar{c}) =  i (\phi - M^{-1}\rho);& \ \ \ \ \ \ \ \  r \phi = 0
\label{roncbara}
\eea
where  
\be
\label{definerho}
\rho \equiv g \ \pi_i^\mathrm{T} \times A_i^\mathrm{T}
\ee
\no is the color-charge density of the transverse degrees of freedom, and we have used
\be
r (M^{-1}\bar{c})  = M^{-1} i D_i \pi_i = i M^{-1}(- D_i \pd_i \phi +g A_i^T \times \pi_i^T).
\ee
Here
\be
\label{definem}
M  \equiv - \pd_i D_i(A^\mathrm{T}) = -  D_i(A^\mathrm{T}) \pd_i
\ee
\no is the Faddeev-Popov operator, and the last equality holds because $A_i^\mathrm{T}$ is transverse.  We see by the preceding definition that $A_i^\mathrm{T}$ and $\pi_i^\mathrm{T}$ are physical, as in the usual Coulomb gauge, whereas $\phi$ is physically equivalent to $M^{-1}\rho$, 
\be
\phi \sim M^{-1}\rho,\nn
\ee
because they differ by an $r$-exact term, and moreover $M^{-1}\rho$ is a function of the transverse degrees of freedom $A_i^\mathrm{T}$ and $\pi_i^\mathrm{T}$.  In the usual Coulomb gauge, the color-Coulomb field has the value $\phi = M^{-1}\rho$.

\section{Dyson-Schwinger equations}\label{sdses}

\begin{figure}
\epsfig{file=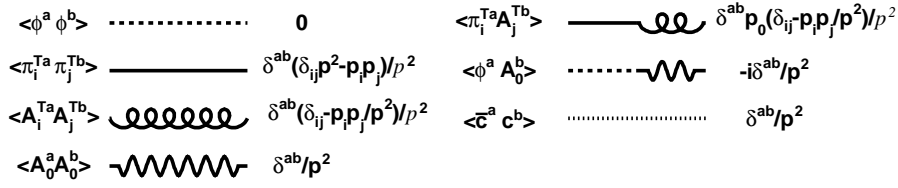,width=\linewidth}
\caption{\label{propfig}The tree-level propagators of all elementary and mixed fields.}
\end{figure}

With the separation $\pi_i = \pi_i^\mathrm{T} - \pd_i\phi$, the on-shell action becomes
\bea
S_{\rm os}  & = & \int d^{s+1}x \Big\{ i \pi_i^\mathrm{T}  \pd_0 A_i^\mathrm{T} + { 1 \over 2} [(\pi^{\rm T})^2 \nn\\
&&+ (\pd_i \phi)^2 + B^2]+ i \pd_i \phi D_i A_0-i g\pi_i^\mathrm{T}  (A_i^\mathrm{T} \times A_0) + \pd_i \bar{c} D_i c  \Big\}.	\label{phasespact}
\eea
\no Transverse propagators are expressed in terms of scalar quantities by
\be
\langle A_i^{{\rm T} a}(x) A_j^{{\rm T} b}(0)\rangle = \delta^{ab} \int d^{s+1}k { \exp(ik \cdot x) \over (2\pi)^{s+1}} P_{ij}(k) D_{AA}^\mathrm{T}(k)\nn
\ee
\no etc., where  $P_{ij}(k) = \delta_{ij} - k_i k_j/{\bf k}^2$ is the projector onto the transverse subspace. The derivation of the DSEs from the action $S_{\rm os}$ is straightforward, keeping in mind that the propagators are 2 by 2 matrices,
\bea
\bpm D_{AA}^\mathrm{T}(k) & D_{A\pi}^\mathrm{T}(k) \cr D_{\pi A}^\mathrm{T}(k) & D_{\pi \pi}^\mathrm{T}(k) \cr \epm,\quad\quad\quad\bpm D_{A_0 A_0}(k) & D_{A_0 \phi}(k) \cr D_{\phi A_0}(k) & D_{\phi \phi}(k) \epm\nn,
\eea
\no that are inverse to 
\bea
\bpm \Gamma_{AA}^\mathrm{T}(k) & \Gamma_{A\pi}^\mathrm{T}(k) \cr \Gamma_{\pi A}^\mathrm{T}(k) & \Gamma_{\pi \pi}^\mathrm{T}(k) \epm,\quad\quad\quad\bpm \Gamma_{A_0A_0}(k) & \Gamma_{A_0 \phi}(k) \cr \Gamma_{\phi A_0}(k) & \Gamma_{\phi \phi}(k) \epm\nn,
\eea
respectively. To define the line-styles for a pictorial representation, the tree-level propagators are shown in figure \ref{propfig}. The tree-level vertices are given in table \ref{tlvtab}.

\begin{table}[h]
\caption{\label{tlvtab}The tree-level 3-point vertices. 4-point vertices will be neglected throughout. Momentum conservation is left implicit. Fields denote the involved fields for the vertex to fix the association of indices, momenta, and field-type.}
\begin{tabular}{|c|c|c|}
\hline
Fields & Notation & Tree-level value \cr
\hline
$A_i^{^\mathrm{T} a}(p) A_j^{^\mathrm{T} b}(q) A_c^{^\mathrm{T} c}(r)$ & $\Gamma^{A^{^\mathrm{T} 3}abc}_{ijk}(p,q,r)$ & $-igf^{abc}((q-r)_i\delta_{jk}+(r-p)_j\delta_{ik}+(p-q)_k\delta_{ij}))$ \cr
\hline
$\phi^a(p) A_i^{^\mathrm{T} b}(q) A_0^c (r)$ & $\Gamma^{\phi A^\mathrm{T} A_0 abc}_{i}(p,q,r)$ & $gf^{abc}p_i$ \cr
\hline
$\pi^{^\mathrm{T} a}_i(p) A_j^{^\mathrm{T} b} (q) A_0^c(r)$ & $\Gamma^{\pi A^\mathrm{T} A_0 abc}_{ij}(p,q,r)$ & $-igf^{abc}\delta_{ij}$ \cr
\hline
$c^a(q) \bar{c}^b(r)A^{^\mathrm{T} c}_i(p)$ & $\Gamma^{c \bar{c} A^T abc}_i(p,q,r)$ & $-igf^{abc} r_i$ \cr 
\hline
\end{tabular}
\end{table}

\begin{figure}
\epsfig{file=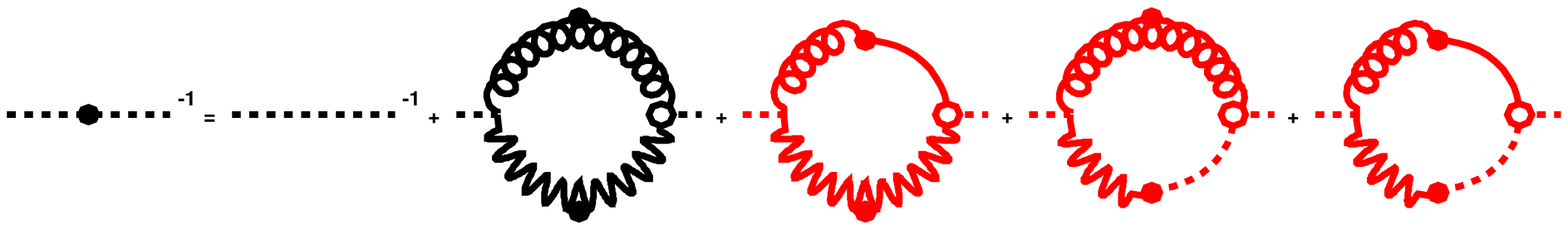,width=\linewidth}\\
\epsfig{file=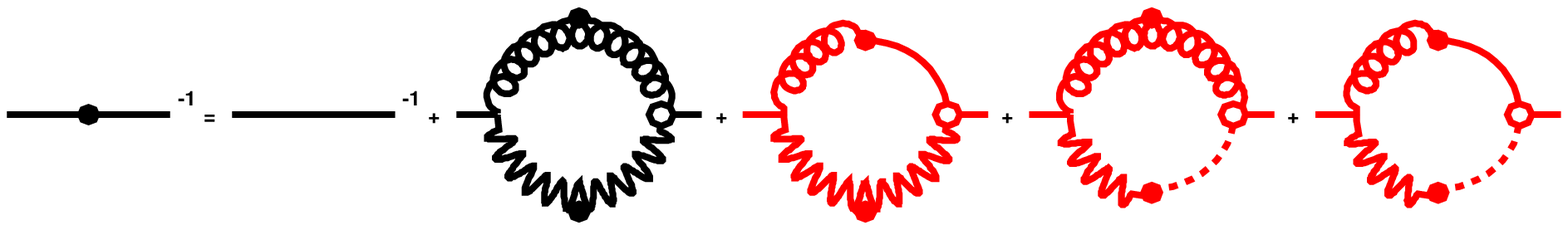,width=\linewidth}\\
\epsfig{file=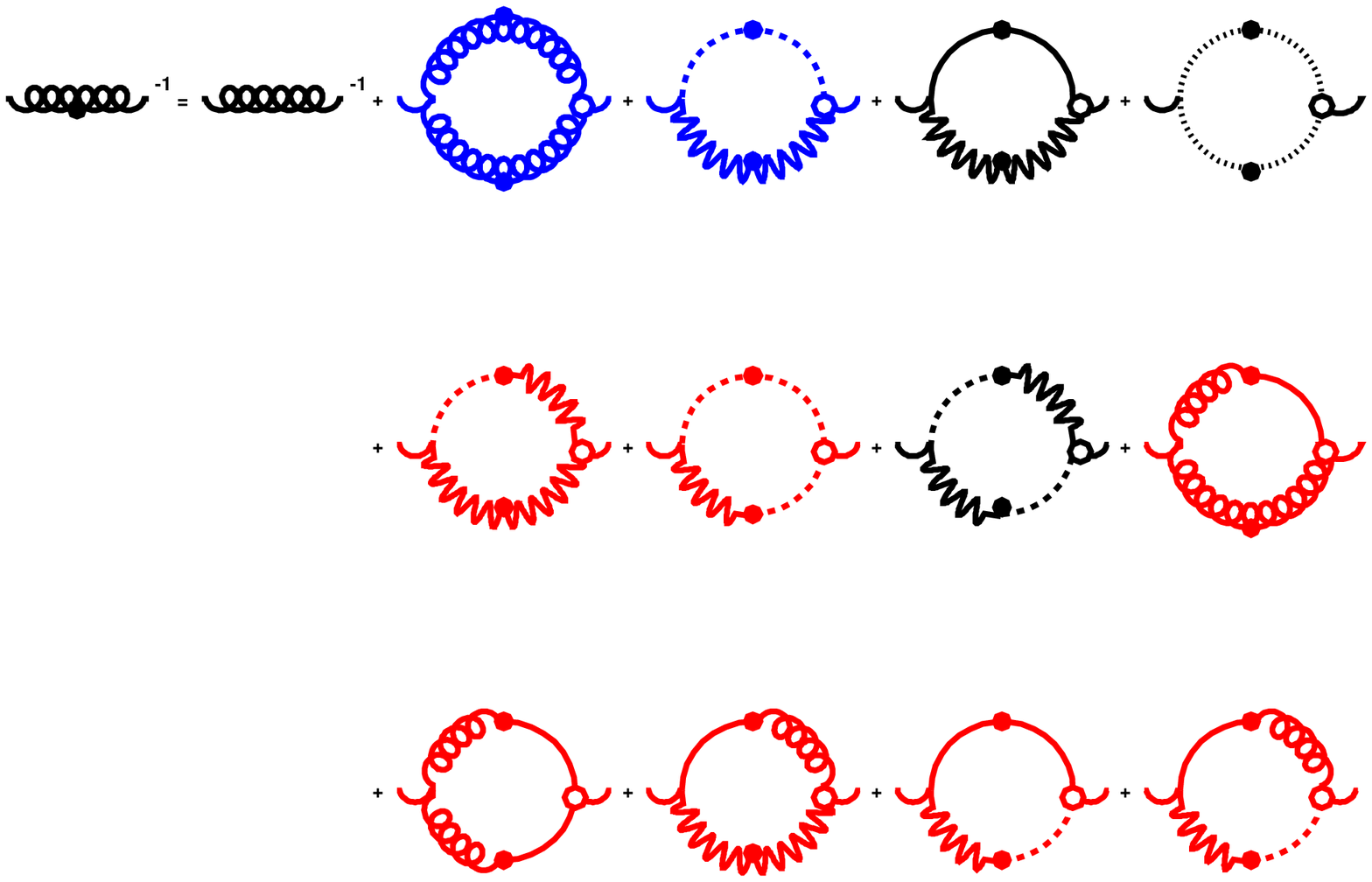,width=\linewidth}\\
\epsfig{file=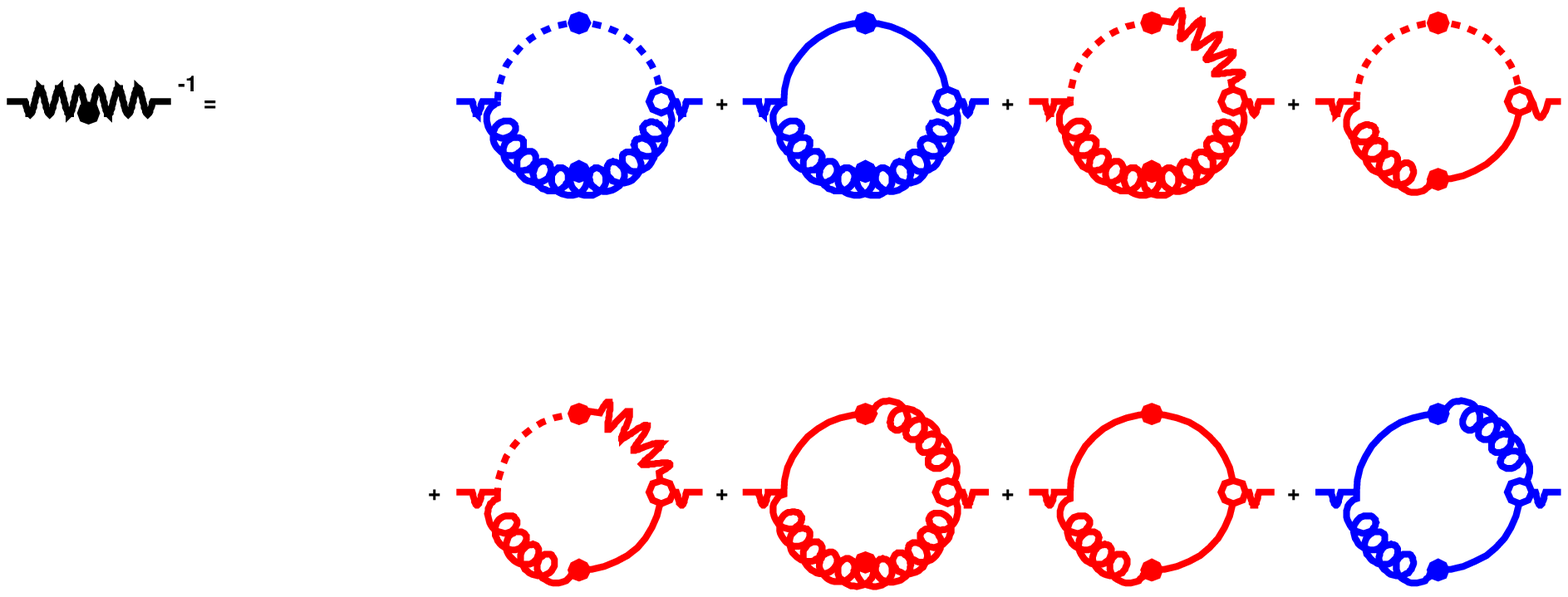,width=\linewidth}
\caption{\label{fphidse}The DSEs for the $\phi$-propagator, $\pi$-propagator, the transverse gluon propagator, and the time-like gluon propagator. Only black diagrams are kept in the truncation. Red diagrams vanish for tree-level vertices. Blue diagrams do not vanish when inserting tree-level vertices, but are discarded in the truncation. Lines with a dot are full propagators, and open circles are full vertices.}
\end{figure}

\begin{figure}
\epsfig{file=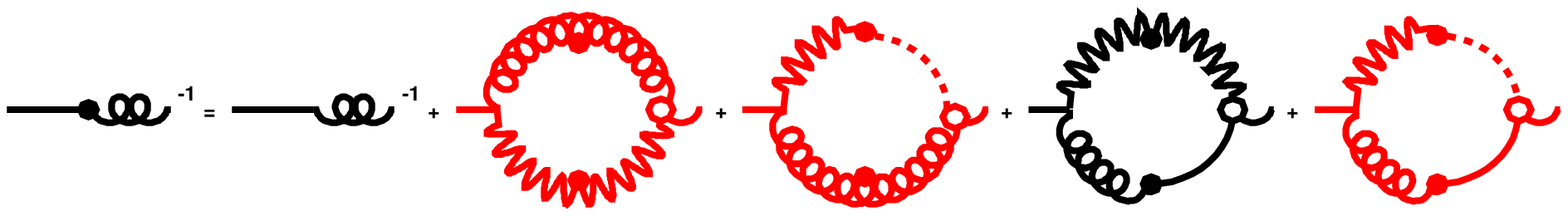,width=\linewidth}\\
\epsfig{file=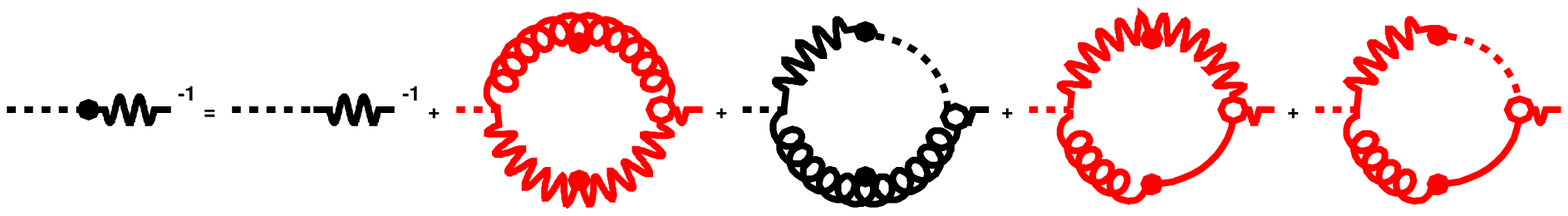,width=\linewidth}\\
\epsfig{file=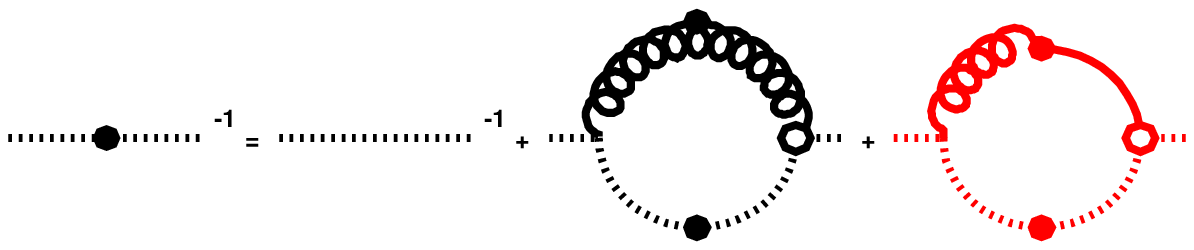,width=\linewidth}
\caption{\label{fpigtdse}The DSE for the mixed $\pi$-transverse-gluon propagator, the mixed $\phi$-$A_0$ propagator, and the ghost propagator. Only black diagrams are kept in the truncation. Red diagrams vanish for tree-level vertices. Lines with a dot are full propagators, and open circles are full vertices.}
\end{figure}

A complete pictorial representation for the Dyson-Schwinger equations, truncated at one-loop level, is given in figures \ref{fphidse} ($\phi$-propagator, $\pi$-propagator, transverse gluon propagator, and the time-like gluon propagator) and \ref{fpigtdse} (mixed $\pi$-transverse gluon propagator, mixed $\phi$-$A_0$ propagator, and ghost propagator). The full expressions will be given only for those contributions present after truncation.  A full set of DSEs could be generated, e.\ g., using the method described in \cite{Alkofer:2008nt}. An independent derivation of these equations can also be found in \cite{watsondses}. No consequences of additional mixed propagators have been included, because even if they are non-zero, they do not survive the proposed truncation. Also, the following expressions for the DSEs will only contain the parts which will be needed explicitly.  The justification for the truncation scheme will be given below.  

The DSE for $\Gamma_{AA}^\mathrm{T}$ reads
\bea
\Gamma^{T ab}_{AA ij}(k)&=&\delta^{ab}\left(\vec{k}^2\delta_{ij}-k_i k_j\right)\nn\\
&&+igf^{adc}\int \frac{d^{s+1}p}{(2\pi)^s} D_{\pi\pi ik}^{Tcf}(p) D_{A_0 A_0}^{de}(k-p)\Gamma^{\pi^\mathrm{T} A^\mathrm{T} A_0 feb}_{kj}(p,k-p,-k)\nn\\
&&+gf^{adc}k_i\int \frac{d^{s+1}p}{(2\pi)^s} D_{A_0\phi}^{cf}(p) D_{A_0\phi}^{de}(k-p)\Gamma_j^{\phi A_0 A^\mathrm{T} feb}(p,k-p,-k)\nn\\
&&+igf^{adc}k_i\int \frac{d^{s+1}p}{(2\pi)^s} D_{c\bar c}^{cf}(p) D_{c\bar c}^{de}(k-p)\Gamma_j^{c\bar c A^\mathrm{T} feb}(p,-k,k-p)\label{ateq}.
\eea
\no and with tree-level vertices
\bea
P_{ij}(k) \ \Gamma_{AA}^\mathrm{T}(k)    &=&  P_{ij}(k) \ {\bf k}^2+ {Ng^2 \over (2\pi)^{s+1}}P_{il}(k) \int d^{s+1}p \Big[P_{lm}(p) \ D_{\pi \pi}^\mathrm{T}(p) \  D_{A_0A_0}(k-p)\nn\\
&&+\  p_l \ D_{A_0 \phi}(p) \  D_{A_0 \phi}(p-k) \ (p-k)_m\nn\\
&&+  p_l \ D_{c \bar{c}}(p) \  D_{c \bar{c}}(p-k) \ (p-k)_m]  \  P_{mj}(k).\nn
\eea
\no $N$ denotes the adjoint Casimir of the gauge group. Upon taking the trace on the spatial indices one obtains the DSE for scalar quantities,
\bea
\Gamma_{AA}^\mathrm{T}(k) &  = & {\bf k}^2+ { Ng^2 \over (2\pi)^{s+1} } \int d^{s+1}p \Big[ {s-2 + (\hat{k} \cdot \hat{p})^2 \over s-1} \ D_{\pi \pi}^\mathrm{T}(p) \  D_{A_0A_0}(k-p)  \label{dsatrans}\\
&&+ {1 - (\hat{k} \cdot \hat{p})^2 \over s-1} \ {\bf p}^2  \ D_{A_0 \phi}(p) \  D_{A_0 \phi}(p-k) + {1 - (\hat{k} \cdot \hat{p})^2 \over s-1} \ {\bf p}^2 \ D_{c \bar{c}}(p) \  D_{c \bar{c}}(p-k)\Big]\nn,
\eea
\no where $\hat{k}$ and $\hat{p}$ denote the corresponding unit vectors. Similarly the equations for $\Gamma_{\pi \pi}^\mathrm{T}$ and $\Gamma_{\pi A}^\mathrm{T}$ read
\bea
\Gamma^{\pi\pi}_{Tij}(k)&=&\delta^{ab}(\delta_{ij}\vec{k}^2+k_i k_j)\label{pieq}\\
&&+igf^{adc}\int \frac{d^{s+1}p}{(2\pi)^s} D_{AA ik}^{T cf}(p) D_{A_0A_0}^{de}(k-p)\Gamma^{\pi^\mathrm{T} A^\mathrm{T} A_0 efb}_{jk}(-k,p,k-p)\nn
\eea
\no or with tree-level vertices
\be
\label{dspiepie}
\Gamma_{\pi \pi}^\mathrm{T}(k)=  1+ { Ng^2 \over (2\pi)^{s+1} } \int d^{s+1}p \  {s-2 + (\hat{k} \cdot \hat{p})^2 \over s-1}  D_{AA}^\mathrm{T}(p) \  D_{A_0A_0}(k-p)  
\ee
\no and 
\be
\label{dspiea}
\Gamma_{\pi A}^\mathrm{T}(k)  =  -k_0+ { Ng^2 \over (2\pi)^{s+1} } \int d^{s+1}p \ {s-2 + (\hat{k} \cdot \hat{p})^2 \over s-1}\ D_{A \pi}^\mathrm{T}(p) \  D_{A_0A_0}(k-p).
\ee
\no For $\Gamma_{A_0A_0}$ we obtain
\be
\label{dsazeroazero}
\Gamma_{A_0A_0}(k) = {Ng^2 \over (2\pi)^{s+1}} \int d^{s+1}p \  \Big[ {\bf k}^2[1 - (\hat{k} \cdot \hat{p})^2] \ D_{AA}^\mathrm{T}(p) \  D_{\phi \phi}(p-k)\Big]. 
\ee
\no The equations \pref{dsazeroazero} and \pref{dspiea} will not be needed in a form with full vertices, and thus these expressions are skipped. For $\Gamma_{\phi \phi}$ we have
\be
\Gamma_{\phi\phi}^{ab}(k)=\delta^{ab}\vec{k}^2+gf^{dae}k_i\int\frac{d^{s+1}p}{(2\pi)^s} D_{AAij}^{Tef}(k-p)D_{A_0A_0}^{dg}(p)\Gamma^{\phi A_0 A^T bgf}_j(-k,p,k-p)\label{phiphieq}.
\ee
\no or with tree-level vertices
\be
\Gamma_{\phi \phi}(k) = {\bf k}^2+ {Ng^2 {\bf k}^2\over (2\pi)^{s+1}} \int d^{s+1}p \ [1 - (\hat{k} \cdot \hat{p})^2] D_{AA}^\mathrm{T}(p) \  D_{A_0 A_0}(p-k)   \ , \nn
\ee
\no and for $\Gamma_{\phi A_0}$,
\be
\Gamma_{\phi A_0}^{ab}(k)=i\delta^{ab} \vec{k}^2+gf^{dae}k_i\int\frac{d^{s+1}p}{(2\pi)^{s+1}} D_{AAij}^{Tef}(k-p)D_{A_0\phi}^{dg}(k)\Gamma^{\phi A_0 A^T bgf}_j(-k,p,k-p)\label{p0eq}\nn
\ee
\no or, with tree-level vertices introduced,
\be
\label{dsphiazero}
\Gamma_{\phi A_0}(k)  = i {\bf k}^2+ {Ng^2{\bf k}^2 \over (2\pi)^{s+1}} \int d^{s+1}p \ [1 - (\hat{k} \cdot \hat{p})^2] D_{AA}^\mathrm{T}(p) \  D_{A_0 \phi}(p-k)   \ . 
\ee
\no Finally for the ghost we obtain
\be
\Gamma_{\bar c c}^{ab}(k)=\delta^{ab} \vec{k}^2-igf^{dae}k_i\int\frac{d^{s+1}p}{(2\pi)^{s+1}} D_{AAij}^{Tef}(k-p)D^{dg}_{c\bar c}(p)\Gamma^{c\bar c A^T bgf}_j(-k,p,k-p)\label{gheq},
\ee
\no which takes for tree-level vertices the form
\be
\label{dsghost}
\Gamma_{\bar{c}  c}(k) =  {\bf k}^2   - {Ng^2{\bf k}^2 \over (2\pi)^{s+1}} \int d^{s+1}p \ [1 - (\hat{k} \cdot \hat{p})^2]D_{AA}^\mathrm{T}(p) \  D_{c \bar{c}}(p-k)   \ , 
\ee
where $\Gamma_{\bar{c}  c}(k) = D_{c \bar{c}}^{-1}(k)$.

\section{Truncation with tree-level vertices}\label{strunc}

Many of the following arguments, and of the truncation, still hold also for the ans\"atze for the vertices which will be made in section \ref{svbt}. For simplicity, here the tree-level expressions will be used, and only commented on, if some argument no longer holds when using non-tree-level vertices.

\subsection{Instantaneous parts of propagators}

In general the propagators depend on two variables, $D(k) = D(|{\bf k}|, k_0)$. However in Coulomb gauge the propagators of the scalar degrees of freedom contain instantaneous parts $D_I$, proportional to $\delta(t)$, and these, in momentum space, depend on a single variable $D_I(|{\bf k}|)$. Here we shall identify the instantaneous parts of propagators.

In the phase space action (\ref{phasespact}), the only time derivative occurs in the term $i \pi_i^\mathrm{T} \pd_0A_i ^\mathrm{T}$.  Accordingly we suppose that the propagators of the transverse degrees of freedom contain no instantaneous parts, but that the scalar fields may have instantaneous parts proportional to $\delta(t)$. It can be shown by appropriately integrating out fields that propagators of the scalar fields may be expressed as
\bea
\delta^{ab} \ D_{c \bar{c}}(x-y) &=& \langle \ (M^{-1})^{ab}(x,y) \  \rangle\nn\\
\delta^{ab} \ iD_{A_0 \phi}(x-y) & = & \langle \ (M^{-1})^{ab}(x,y) - \  (K\rho)^a(x) \  (M^{-1}\rho)^b(y) \ \rangle\nn\\
\delta^{ab} \ D_{A_0 A_0}(x-y)   & = & \langle \ K^{ab}(x,y) -  \ (K\rho)^a(x) \ (K\rho)^b(y) \  \rangle\nn\\
\delta^{ab} \ D_{\phi \phi}(x-y) &  = &\langle \ (M^{-1}\rho)^a(x) \ (M^{-1}\rho)^b(y) \ \rangle,\nn
\eea
\no where $K$ is the operator with kernel
\be
K^{ab}(x,y) \equiv [M^{-1} (- \nabla^2) M^{-1}]^{ab}(x,y).
\ee
We have $\rho = g \pi_i^\mathrm{T} \times A_i^\mathrm{T}$ and $M = - \pd_i D_i(A^\mathrm{T})$ so the quantities on the right depend only on the transverse degrees of freedom.  The propagators of the scalar fields contain instantaneous parts because the kernel $M^{-1}(x,y) = M^{-1}({\bf x}, {\bf y}) \delta(x_0 - y_0)$ is instantaneous, and consequently so is $K(x,y) = K({\bf x}, {\bf y}) \delta(x_0 - y_0)$.

Assuming that the transverse fields do not contribute to the instantaneous parts, the last equations give the decomposition of the propagators into instantaneous~(I) and non-instantaneous~(N) parts, which reads in momentum space
\bea
D_{c \bar{c}}(k) &=& D_{I c \bar{c}}(|{\bf k}|)\nn\\
D_{A_0 \phi}(k) &=& D_{IA_0 \phi}(|{\bf k}|) + D_{NA_0 \phi}(k)\nn\\
D_{A_0 A_0}(k)  &  = &D_{IA_0 A_0}(|{\bf k}|) + D_{N A_0 A_0}(k)\nn\\
D_{\phi \phi}(k) & =& D_{N\phi \phi}(k).\nn
\eea
Moreover we have the equality of the instantaneous Bose and Fermi propagators   
\be
i D_{IA_0 \phi}(|{\bf k}|) = D_{I c \bar{c}}(|{\bf k}|).
\ee
which may be shown to be a consequence of the $r$-symmetry.

The ghost propagator $D_{c \bar{c}}(|{\bf k}|)$ is purely instantaneous. The integral over $p_0$ in (\ref{dsghost}) yields
\be
{1 \over 2\pi} \int dp_0 \ D_{AA}^\mathrm{T}(p) = D_{=AA}(|{\bf p}|),
\ee
where $D_{=AA}(|{\bf k}|)$ is the equal-time $A^\mathrm{T}$-$A^\mathrm{T}$ propagator in momentum space.  The DSE of the ghost, 
\be
D_{c \bar{c}}^{-1}(|{\bf k}|)  = {\bf k}^2- {Ng^2{\bf k}^2 \over (2\pi)^s} \int d^sp \ [1 - (\hat{k} \cdot \hat{p})^2]D_{=AA}(|{\bf p}|) \  D_{c \bar{c}}(|{\bf p-k}|), 
\label{dsghostinst}
\ee
thus involves unknown functions of only one variable.

Later on, in section \ref{svbt}, we will presume that the momentum dependence of the vertex dressing functions appearing also are purely instantaneous, i.\ e., only depend on $\vec{k}$ and not on $k_0$. Therefore, everything in this subsection also applies to this case.

\subsection{Instantaneous approximation and infrared limit}

We wish to explore the hypothesis that there exists an asymptotic infrared limit of the DSEs which is dominated by loops containing an instantaneous propagator. As Ansatz we suppose that we may neglect loops that contain no instantaneous propagators on the RHS of the DSEs for scalar propagators, eqs.~(\ref{dsazeroazero}) -- (\ref{dsphiazero}).  This is equivalent to the substitutions on the RHS
\bea
D_{\phi \phi}(|{\bf k}|, k_0) &\rightarrow &D_{I \phi \phi}(|{\bf k}|) = 0  \nonumber \\
D_{A_0 A_0}(|{\bf k}|, k_0) &\rightarrow &D_{I A_0 A_0}(|{\bf k}|)   \nonumber \\
iD_{A_0 \phi}(|{\bf k}|, k_0) &\rightarrow &iD_{I A_0 \phi}(|{\bf k}|) = D_{c \bar{c}}(|{\bf k}|).
\eea
\no Equations~(\ref{dsazeroazero}) -- (\ref{dsphiazero}) then yield on the LHS
\bea
\Gamma_{A_0 A_0} &=& 0\nn\\
- i\Gamma_{I \phi A_0} &=& \Gamma_{ \bar{c} c}\nn\\
\Gamma_{\phi \phi}(|{\bf k}|) &  = & {\bf k}^2+ {Ng^2{\bf k}^2 \over (2\pi)^s} \int d^sp \ [1 - (\hat{k} \cdot \hat{p})^2]D_{=AA}(|{\bf p}|) \  D_{IA_0 A_0}(|{\bf p-k}|) , \label{dsazeroinst}
\eea
\no which are all consistent with our Ansatz.
We have
\be
D_{IA_0 A_0}(|{\bf k}|) = {- \Gamma_{\phi \phi}(|{\bf k}|)     \over  \Gamma_{A_0 \phi}^2(|{\bf k}| )}
\ee
\no which gives
\be
\Gamma_{\phi \phi}(|{\bf k}|) = { D_{IA_0 A_0}(|{\bf k}|)    \over  D_{c \bar{c}}^2(|{\bf k}| )}.
\ee
\no This implies
\be
{ D_{IA_0 A_0}(|{\bf k}|)    \over  D_{c \bar{c}}^2(|{\bf k}| )}     = {\bf k}^2   + {Ng^2{\bf k}^2 \over (2\pi)^s} \int d^sp \ [1 - (\hat{k} \cdot \hat{p})^2] D_{=AA}(|{\bf p}|) \  D_{IA_0 A_0}(|{\bf p-k}|).
\label{dsdazeroazero}
\ee
\no The pair of equations (\ref{dsghostinst}) and (\ref{dsdazeroazero}) determine the two propagators $D_{c \bar{c}}( |{\bf k}| )$ and $D_{IA_0 A_0}(|{\bf k}|)$, provided that $D_{=A A}(|{\bf k}|)$ is known. 

We now consider the DSEs for the transverse propagators.  We make the same approximation $D_{A_0 A_0}(k) \rightarrow D_{IA_0 A_0}(|{\bf k}|)$ into the RHS of (\ref{dspiea}) which yields the integral
\be
\int dp_0 \ D_{A \pi}^\mathrm{T}(p) = 0,
\ee
which vanishes in the present approximation because $D_{A \pi}^\mathrm{T}(p_0, |{\bf p}|)$ is odd in $p_0$, as one sees from the tree-level term in (\ref{dspiea}), and as will be verified below. This gives
\be
\label{dspieainst}
\Gamma_{\pi A}^\mathrm{T}(|{\bf k}|) = -k_0.
\ee
We make the same substitution $D_{A_0 A_0} \rightarrow D_{IA_0 A_0}$ in (\ref{dspiepie}) and obtain
\be
\Gamma_{\pi \pi}^\mathrm{T}(|{\bf k}|)  =  1+ {Ng^2 \over (2\pi)^s} \int d^sp \ { s-2 + (\hat{k} \cdot \hat{p})^2 \over s-1}D_{=AA}(|{\bf p}|) \  D_{IA_0A_0}(|{\bf k-p}|).\label{dspiepieinst}
\ee
There is no $k_0$ dependence on the RHS, so, as indicated, $\Gamma_{\pi \pi}^\mathrm{T}(|{\bf k}|)$ depends on the one variable $|{\bf k}|$.

There remains to consider the DSE for $\Gamma^\mathrm{T}_{AA}$.  Upon making the substitution $iD_{A_0 \phi} \rightarrow D_{c \bar{c}}$ into the RHS of (\ref{dsatrans}) in the loops that contain instantaneous pieces, one finds that the loop of scalar bosons $D_{IA_0 \phi}(p)D_{IA_0 \phi}(p-k)$ exactly cancels the ghost loop $D_{c \bar{c}}(p)D_{c \bar{c}}(p-k)$. This is a crucial cancellation, and it holds in this truncation only with tree-level vertices. Beyond such vertices, this is non-trivial, and implies constraints which will be discussed in section \ref{svbt}. Of course, in the full solution this would be guaranteed by the $r$ symmetry.

This leaves only one loop in the equation, and we obtain 
\be
\Gamma_{AA}^\mathrm{T}(|{\bf k}|)   = {\bf k}^2+ {Ng^2 \over (2\pi)^s} \int d^sp \ { s-2 + (\hat{k} \cdot \hat{p})^2 \over s-1}D_{=\pi \pi}(|{\bf p}|) \  D_{IA_0A_0}(|{\bf k - p}|) . \label{dsatransinst}
\ee
Again there is no $k_0$ dependence on the RHS, so $\Gamma_{AA}^\mathrm{T}(|{\bf k}|)$ is a function of $|{\bf k}|$ only.

It should be noted that there is a certain ambiguity involved here: In the DSE (\ref{dsatrans}) for $\Gamma^\mathrm{T}_{AA}$ there is a cross term 
$D_{IA_0 \phi}(p)D_{NA_0 \phi}(p-k)$  containing an instantaneous propagator (I) multiplying a non-instantaneous propagator (N), which in principle should be kept.  However, if we keep the one instantaneous part on the RHS of the DSE (\ref{dsphiazero}) for $\Gamma_{\phi A_0}$, then $\Gamma_{\phi A_0}$ comes out purely instantaneous, so within our approximation  $D_{N \phi A_0} = 0$, and we have no way to calculate a non-zero value for this quantity.  This is self-consistent. However, this assumption may be one of the reasons for the failure of finding a solution with the desired properties in this truncation.

Using the results of the last three equations, we see that the transverse $\Gamma$-matrix is given by
\be
\bpm \Gamma_{AA}^\mathrm{T}(|{\bf k}|) & k_0 \cr - k_0  & \Gamma_{\pi \pi}^\mathrm{T}(|{\bf k}|) \cr \epm\nn
\ee
\no The elements of the inverse matrix are thus
\bea
D_{AA}^\mathrm{T}(k_0, |{\bf k}|) & = & \Gamma_{\pi \pi}^\mathrm{T}(|{\bf k}|)/ Q   \nonumber  \\
D_{A\pi}^\mathrm{T}(k_0, |{\bf k}|) & = & - k_0 / Q   \nonumber  \\
D_{\pi \pi}^\mathrm{T}(k_0, |{\bf k}|) & = & \Gamma_{AA}^\mathrm{T}(|{\bf k}|)/ Q,
\eea
\no where
\be
Q = k_0^2 + \Gamma_{AA}^\mathrm{T}(|{\bf k}|) \ \Gamma_{\pi \pi}^\mathrm{T}(|{\bf k}|).
\ee
\no We see that $D_{A\pi}^\mathrm{T}(k_0, |{\bf k}|)$ is odd in $k_0$, as asserted.

It follows that the equal-time propagators that appear on the RHS of the DS equations are given by
\bea
D_{=AA}(|{\bf p}|) & = & {1 \over 2\pi} \int dp_0 \ D_{AA}^\mathrm{T}(p)\nonumber \\ 
& = & {1 \over 2\pi} \int dp_0 \ { \Gamma_{\pi \pi}^\mathrm{T}(|{\bf p}|) \over  p_0^2 + \Gamma_{AA}^\mathrm{T}(|{\bf p}|) \ \Gamma_{\pi \pi}^\mathrm{T}(|{\bf p}|)} ={1\over 2} \Big[{\Gamma_{\pi \pi}^\mathrm{T}(|{\bf p}|) \over \Gamma_{A A}^\mathrm{T}(|{\bf p}|)}\Big]^{1/2},\nn
\eea
\no and similarly
\be
D_{=\pi \pi}(|{\bf p}|)= {1\over 2} \Big[{\Gamma_{A A}^\mathrm{T}(|{\bf p}|) \over \Gamma_{\pi \pi}^\mathrm{T}(|{\bf p}|)}\Big]^{1/2}.
\ee
\no This implies that
\be
D_{=\pi \pi}(|{\bf p}|) = {1 \over 4} D_{=AA}^{-1}(|{\bf p}|),
\ee
\no so there is only one unknown function of one variable.  Upon taking the ratio of (\ref{dspiepieinst}) and (\ref{dsatransinst}) we obtain
\be
4 D_{=AA}^2(|{\bf k}|) = { \Gamma_{=\pi \pi}(|{\bf k}|) \over \Gamma_{=AA}(|{\bf k}|) } = {X \over Y},
\ee
\no where $X$ and $Y$ are the RHS of (\ref{dspiepieinst}) and (\ref{dsatransinst}) respectively.  This provides the desired equation for $D_{=AA}(|{\bf k}|)$.  However we expect that $D_{IA_0A_0}(|{\bf p-k}|)$ is highly singular at ${\bf k} = {\bf p}$, for example like $1 / |{\bf p-k}|^4$, so there are infrared divergences in $X$ and $Y$.  However, these cancel by rearranging the last equation to read 
\bea
&&4 {\bf k}^2 D_{=AA}(|{\bf k}|) - D_{=AA}^{-1}(|{\bf k}|)   \label{cancelir}\\
&& = {Ng^2 \over (2\pi)^s} \int d^sp \ {  s-2 + (\hat{p} \cdot \hat{k})^2 \over s-1}  \ D_{IA_0 A_0}(|{\bf k - p}|)\Big[ { D_{=AA}(|{\bf p}|) \over D_{=AA}(|{\bf k}|) } - { D_{=AA}(|{\bf k}|) \over D_{=AA}(|{\bf p}|) } \Big].\nonumber
\eea
In all we have three equations, (\ref{dsghostinst}) and (\ref{dsdazeroazero}) and (\ref{cancelir}) for the three functions $D_{c \bar{c}}(|{\bf p}|)$, $D_{IA_0 A_0}(|{\bf p}|)$ and $D_{=AA}(|{\bf p}|)$.

\subsection{Infrared asymptotic solution}

We suppose that these three propagators have infrared asymptotic limits that are described by power laws,
\bea
\label{powerlaws}
D_{c \bar{c}}(|{\bf p}|)& =& { b_\Delta \over ({\bf p}^2)^{1 + \kappa_\Delta} }\nonumber  \\
g^2D_{I A_0 A_0}(|{\bf p}|)& =& { b_V \over ({\bf p}^2)^{1 + \kappa_V} }\nonumber  \\
g^2 D_{=AA}(|{\bf p}|)& =& { b_T \over ({\bf p}^2)^{1 + \kappa_T} }.
\eea  

\subsubsection{Solution of DSE for $D_{c \bar{c}}$}

As discussed elsewhere \cite{Zwanziger:1998ez,Gribov:1977wm} we choose a solution for which the ghost propagator $D_{c \bar{c}}(|{\bf k}|)$ is more singular than the free propagator $1/{\bf k}^2$ at ${\bf k} = 0$, implementing the horizon condition.  We therefore require that the term quadratic in $\bf k$ on the r.h.s.\ of (\ref{dsghostinst}) is canceled, which gives
\be
D_{c \bar{c}}^{-1}(|{\bf k}|)   =  {Ng^2{\bf k}^2 \over (2\pi)^s} \int d^sp \ [1 - (\hat{k} \cdot \hat{p})^2] \ D_{=AA}(|{\bf p}|)[D_{c \bar{c}}(|{\bf p}|)- D_{c \bar{c}}(|{\bf p-k}|)].\label{dsghosthor}
\ee
This choice of a cancellation mechanism makes the equation invariant under the renormalization group (RG), as it must be.

By comparing powers of momenta on the left and right one finds from the power laws (\ref{powerlaws}) that the critical exponents satisfy
\be
\label{relatekappas}
\kappa_T + 2 \kappa_\Delta = (s - 4)/2.
\ee
The integral is ultraviolet convergent provided that $s/2 < 3+\kappa_T + \kappa_\Delta$, and infrared convergent provided that $s/2 > 2 + \kappa_T + \kappa_\Delta$, which gives
\be
\label{boundk}
0 < \kappa_\Delta < 1.
\ee
\no The integral on the r.h.s.\ of (\ref{dsghosthor}) was evaluated in eq.~(A.17) of \cite{Zwanziger:2001kw}, with the result\footnote{Due to the structural equivalence to Landau gauge it seems likely that a more general analysis, like in the Landau gauge case \cite{Watson:2001yv}, is possible.}
\be
(b_T b_\Delta^2)^{-1}  =  { N(s-1) \ \Gamma(1 - \kappa_\Delta) \ \Gamma(2 \kappa_\Delta) \over (4 \pi)^{s/2} \ \Gamma(1 + \kappa_\Delta) \ \Gamma(s/2 - 2\kappa_\Delta)  }{   \Gamma(s/2 - \kappa_\Delta) \over \ \Gamma(1 + s/2 + \kappa_\Delta)  }\label{Ione}
\ee
where we have used $\pi / \sin(\pi \kappa_\Delta) = - \Gamma(- \kappa_\Delta) \Gamma(1 + \kappa_\Delta)$.

\subsubsection{Solution of DSE for $D_{IA_0 A_0}$}

We next substitute the power laws (\ref{powerlaws}) into (\ref{dsdazeroazero}).  The integral is ultraviolet convergent provided that
$s/2 < 2 + \kappa_T + \kappa_V$, and is infrared convergent provided that $s/2 > \kappa_V$.  This gives
\be
\label{convIphi}
2 \kappa_\Delta < \kappa_V < s/2,
\ee 
\no where we have again used~(\ref{relatekappas}).  The l.h.s.\ of (\ref{dsdazeroazero}) is given by $b_V b_\Delta^{-2} ({\bf k}^2)^{1 + 2\kappa_\Delta - \kappa_V}$.  By the last inequality, this dominates ${\bf k}^2$ at low $\bf k$, and we conclude that the tree-level term in (\ref{dsdazeroazero}) is negligible in the infrared, and (\ref{dsdazeroazero}) simplifies to
\be
{ D_{IA_0 A_0}(|{\bf k}|)     \over  D_{c \bar{c}}^2(|{\bf k}| )}      =     {Ng^2{\bf k}^2 \over (2\pi)^s} \int d^sp \ [1 - (\hat{k} \cdot \hat{p})^2] D_{=AA}(|{\bf p}|) \  D_{IA_0 A_0}(|{\bf p-k}|).\label{dsdazeroazeroh}
\ee
\no This equation is also invariant under RG transformations.

We substitute the power laws (\ref{powerlaws}) into this equation, and upon equating powers of momenta, we again obtain~(\ref{relatekappas}).
It is straightforward to evaluate this integral which yields, 
\be
(b_T b_\Delta^2)^{-1}  =  { N(s-1) \ \Gamma(s/2 - \kappa_V) \ \Gamma(1 + 2 \kappa_\Delta)\over 2(4 \pi)^{s/2} \ \Gamma(s/2 - 2 \kappa_\Delta) \ \Gamma(1 + \kappa_V)  }{  \ \Gamma(\kappa_V - 2\kappa_\Delta)  \over \ \Gamma(1 + s/2 - \kappa_V + 2\kappa_\Delta)  },
\label{Itwo}
\ee
\no where we used~(\ref{relatekappas}) to eliminate $\kappa_T$.  Upon comparing with (\ref{Ione}), we obtain a relation between $\kappa_\Delta$ and $\kappa_V$,
\be
{ \kappa_\Delta \  \Gamma(s/2 - \kappa_V)  \  \Gamma(\kappa_V - 2\kappa_\Delta)\over   \ \Gamma(1 + \kappa_V)   \ \Gamma(1 + s/2 - \kappa_V + 2\kappa_\Delta) } = {    \Gamma(1 - \kappa_\Delta)\ \Gamma(s/2 - \kappa_\Delta)  \over\Gamma(1 + \kappa_\Delta) \ \Gamma(1 + s/2 + \kappa_\Delta)  }.
\label{Vanddelta} 
\ee

\subsubsection{Solution of the DSE for $D_{=AA}^\mathrm{T}$}

Finally we consider eq.\ (\ref{cancelir}) for the transverse propagator. We shall show that at small ${\bf k}$ the loop integral dominates the tree-level terms.  The tree-level terms on the l.h.s.\ are of order $|{\bf k}|^{ - 2 \kappa_T} = |{\bf k}|^{4\kappa_\Delta + 4 - s}$ and 
$|{\bf k}|^{2 + 2 \kappa_T} = |{\bf k}|^{ s - 2 - 4\kappa_\Delta}$ respectively.  The ratio of these terms is $|{\bf k}|^{8\kappa_\Delta + 6 - 2s}$.  From the inequality (\ref{boundk}) we have $\kappa_\Delta > 0$, so this power is positive for $s \leq 3$.  Thus, at small momentum the second tree-level term $|{\bf k}|^{ s - 2 - 4\kappa_\Delta}$ dominates the first.  Moreover the dimension of the r.h.s.\ is $|{\bf k}|^{s - 2 - 2\kappa_V}$, so it, in turn dominates the second tree-level term by the inequality (\ref{convIphi}), $\kappa_V > 2\kappa_\Delta$.  We conclude that the loop integral in (\ref{cancelir}) dominates both tree-level terms in the infrared asymptotic limit, and this equation simplifies to
\be
0  =  {Ng^2 \over (2\pi)^s} \int d^sp \ [ s-2 + (\hat{p} \cdot \hat{k})^2] \ D_{IA_0 A_0}(|{\bf k - p}|)\Big[ { D_{=AA}(|{\bf p}|) \over D_{=AA}(|{\bf k}|) } - { D_{=AA}(|{\bf k}|) \over D_{=AA}(|{\bf p}|) } \Big].
\label{irtransvers}
\ee
\no This equation is also RG-invariant.

By inspection one sees that this equation is satisfied when $D_{=AA}(|{\bf k}|) = D_{=AA}(|{\bf p}|)$, which occurs at
\be 
\kappa_T = -1,
\ee
or equivalently, at
\be
\label{solvekd}
\kappa_\Delta = (s - 2)/4.
\ee
\no For this value of $\kappa_\Delta$, the bounds (\ref{convIphi}) read
\be
\label{convIphia}
(s - 2)/2 < \kappa_V < s/2.
\ee	
\no This expression for $\kappa_\Delta$ vanishes for $s = 2$, and is negative for $s < 2$, but by the inequality (\ref{boundk}) we have
$\kappa_\Delta > 0$  (This expression also violates the horizon condition for $s \leq 2$.).  So the obvious solution to (\ref{irtransvers}), namely $\kappa_T = -1$, does not give a consistent solution for $s \leq 2$. 

\subsection{Determination of critical exponents}\label{scrit}

We substitute the value just obtained, $\kappa_\Delta = (s-2)/4$, into (\ref{Vanddelta}) and obtain an equation for $\kappa_V$,
\be
\label{solveV} 
{ ({s-2 \over 4}) \  \Gamma(s/2 - \kappa_V)   \  \Gamma(\kappa_V - s/2 + 1) \over    \Gamma(1 + \kappa_V)  \ \Gamma(s - \kappa_V) }  = {    \Gamma( {6-s \over 4} ) \over \Gamma( {3s+2 \over 4} )   }.
\ee
\no To simplify this equation, we use the identity
\be
\Gamma(s/2 - \kappa_V)   \  \Gamma(\kappa_V - s/2 + 1) = { \pi \over \sin[\pi(s/2 - \kappa_V)] }   = { \pi \over \cos(\pi \kappa'_V) }.\nn
\ee
\no Here the shifted variable $\kappa'_V$ is defined by
\be
\kappa'_V \equiv \kappa_V - (s-1)/2,
\ee
\no and the bounds (\ref{convIphia}) read
\be
\label{boundkprime}
-1/2  < \kappa'_V < 1/2.
\ee
\no The shifted variable $\kappa'_V$ is determined by
\be
{\cos(\pi \kappa'_V) \  \Gamma[ ({s+1 \over 2}) + \kappa'_V ] \ \Gamma[ ({s+1 \over 2}) - \kappa'_V ] \over \pi}= { ({s-2 \over 4}) \   \Gamma( {3s+2 \over 4} )  \over\Gamma( {6-s \over 4} )  }.\label{solvekv}
\ee
\no The l.h.s.\ is symmetric about $\kappa'_V = 0$, so if $\kappa'_V$ is a solution so is $-\kappa'_V$. Note that $\kappa'_V = 0$ is equivalent to $\kappa_V = {s-1 \over 2}$, or $g^2 D_{IA_0 A_0} = { b_V \over ({\bf k}^2)^{(s+1)/2} }$.  This describes  a linearly rising color-Coulomb potential in spatial dimension $s$.

Our results are not unreasonable at the physical value $s = 3$.  Indeed, at $s = 3$, the r.h.s.\ of the last equation has the value 
${ (1/4) \Gamma(11/4) \over \Gamma(3/4) } = 21/64 \approx 0.328$, while in the allowed interval (\ref{boundkprime}), the l.h.s.\ of (\ref{solvekv}) has a maximum at $\kappa'_V= 0$ where it has the value $1/\pi \approx 0.318$.  Thus for $s = 3$
the two sides agree to within 3\%, at $\kappa'_V = 0$, which describes a linearly rising color-Coulomb potential, in accordance with numerical simulations \cite{Greensite:2003xf}. 

Note, however, that the maximum of the l.h.s.\ of (\ref{solvekv}) in the allowed interval is $1/\pi$, and $1/\pi < 21/64$. Hence, strictly speaking, there is no real solution at $s = 3$.  As $s$ decreases from $s = 3$,  a real solution develops at $\kappa'_V = 0$ and $s = s_c$, where $s_c$ is a critical dimension that satisfies~(\ref{solvekv}) at $\kappa'_V = 0$,
\be
\label{solvesc}
{ \Gamma^2({s+1 \over 2}) \over \pi}= { ({s-2 \over 4}) \   \Gamma( {3s+2 \over 4} )  \over\Gamma( {6-s \over 4} )  },
\ee
\no namely
\be
s_c \approx 2.9665.
\ee
\no Recall that $\kappa'_V = 0$ is equivalent to $\kappa_V = (s_c-1)/2$, and this corresponds to  a linearly rising color-Coulomb potential at $s = s_c$.  As $s$ decreases below the critical dimension $s_c$, the solution $\kappa'_V = 0$ bifurcates into two solutions that approach $\kappa'_V = \pm 1/2$, as $s$ approaches $s = 2$. Although neither of the two branches corresponds to a linearly rising color-Coulomb potential, it is intriguing that the average of the two solutions $\pm \kappa'_V$ is $\kappa'_V = 0$, or $\kappa_V = (s-1)/2$, and this describes a linearly rising potential for all $s$.

We have seen that there is no consistent solution for $s < 2$, and in fact the r.h.s.\ of (\ref{solvekv}) is negative for $s < 2$, whereas the l.h.s.\ is positive for $\kappa'_V$ in the allowed interval~(\ref{boundkprime}). The question, whether vertex corrections could provide a consistent solution will be discussed in section \ref{svbt}. Unfortunately, it turns out that this is not the case, at least, at the next-to-trivial level.

	Despite the near success in $s = 3$ spatial dimensions, we consider the failure at $s = 2$ to indicate a serious flaw of the present truncation scheme.

\subsection{Infrared fixed point of running coupling constant}

In \cite{Fischer2005}, a running coupling constant was defined in gauges that interpolate between the Landau gauge and the Coulomb gauge.  It was found that in Landau gauge and in all interpolating gauges this running coupling has an infrared fixed point 
\be
\alpha_{\rm interpolating}(0) = {8.915 \over N}.	
\ee
\no In Coulomb gauge this infrared fixed point is given~\cite{Fischer2005} by the formula (in present notation)
\be
\alpha_{\rm coulomb}(0) = {4 b_\Delta^2 b_T \over 3 \pi}.
\ee
\no However the Coulomb gauge is a singular limit of the interpolating gauge, and the methods used in \cite{Fischer2005} were not sufficient to determine the numerical value of the infrared fixed point in Coulomb gauge. However our (near) solution of the Coulomb-gauge DSE provides a value. We take spatial dimension $s = 3$, for which $\kappa_\Delta$ in (\ref{solvekd}) has the value $\kappa_\Delta = 1/4$, and substitute these values into (\ref{Ione}). This gives
\be
(b_T b_\Delta^2)^{-1} = {4N \over 21 \pi},
\ee
\no and we obtain
\be
\alpha_{\rm coulomb}(0) = {7 \over N}.
\ee
It is not clear what is the origin of the difference between interpolating gauges and the Coulomb gauge. There could be a discontinuity but, if so, the discontinuity is not very large. Note, however, that different approximations are made in the two calculations. Indeed here we used the phase space representation for the Coulomb gauge, whereas in interpolating gauges the configuration space representation was used, so the truncation schemes are not identical even though both use tree-level vertices. Furthermore, lattice investigations of the Coulomb gauge limit of the interpolating gauge suggest also a smooth limit, except at zero momentum \cite{Cucchieri:2007uj}.

\section{Vertices beyond tree-level}\label{svbt}

Since the truncation proposed ultimately has failed, it is necessary to reconsider it. The simplest way is to include vertex correction, and check, whether they can be arranged such as to arrive at a consistent solution\footnote{Note that a similar program has also bee pursued in Landau gauge in \cite{Atkinson:1998zc}.}. The most primitive vertex modification in this truncation is to keep the tree-level Lorentz structure, but to include a scalar dressing function for each vertex appearing. In particular, these corrections will also be of instantaneous type. This implies that only the DSEs \pref{cancelir}, \pref{dsghosthor}, and \pref{dsdazeroazeroh} are affected, as all the other consequences of the truncation remain valid.

Except for one important caveat. One of the central elements in the truncation so far was the cancellation of the ghost loop and the $\phi A_0$-loop in the DSE for the transverse gluon propagator. That is true at one-loop order in perturbative calculations, since the propagators are identical, apart from a factor $i$. This identity persists beyond one-loop perturbation theory also in the instantaneous approximation employed here. However, this alone does not guarantee the cancellation of the loops. Either the vertices must also coincide, or their differences must arrange, in a subtle way, to allow the cancellation, at least, for the purpose at hand, the cancellation of the leading infrared part. To achieve self-consistency, the cancellation has therefore to be built in into the ans\"atze for the vertices.

Therefore, the DSEs to be satisfied are a system of equations for the three unknown functions
\be
D_{=A^\mathrm{T} A^\mathrm{T}}\quad\quad D_G\quad\quad D_{00}\nn
\ee
\no and in addition also have to act as constraint equations which restrict the form of possible vertex modifications for the three vertices
\be
\Gamma^{A^\mathrm{T}\bar c c}\quad\quad\Gamma^{A^\mathrm{T} A_0\phi}\quad\quad\Gamma^{A^\mathrm{T} A_0\pi^\mathrm{T}}.\nn
\ee
\no The calculations in section \ref{strunc} to obtain the DSEs \pref{cancelir}, \pref{dsghosthor}, and \pref{dsdazeroazeroh} can still be made in the same way, it is only necessary to keep the scalar dressings of the vertices at each step.

To solve the equations, the same ans\"atze \pref{powerlaws} for the propagators will be made. For the scalar vertex dressings the ans\"atze
\bea
\Gamma^{A^\mathrm{T}\bar c c}(p,q,k)&=&a_{\alpha}p^{2\alpha_T}q^{2\alpha_\Delta}k^{2\alpha_\Delta}\nn\\
\Gamma^{A^\mathrm{T} A_0\phi}(p,q,k)&=&a_{\alpha}p^{2\alpha_T}q^{2\alpha_\Delta}k^{2\alpha_\Delta}\nn\\
\Gamma^{A^\mathrm{T} A_0\pi^\mathrm{T}}(p,q,k)&=&a_{\beta}p^{2\beta_T}q^{2\beta_0}k^{2\beta_\pi}\nn
\eea
\no will be made. Herein, anticipating a ghost-anti-ghost symmetry as in Landau gauge \cite{Alkofer:2000wg}, the exponents for the ghost- and anti-ghost-leg of the ghost-gluon vertex have been set equal. Furthermore, to guarantee exact cancellation of the ghost diagram in the transverse gluon equation by the $A_0\phi$ diagram, the parameters of the ghost-gluon vertex and the $A^\mathrm{T} A_0\phi$ have been set equal\footnote{If this constraint is not included, it is possible to construct a solution with $\kappa_V=1$ with a bare ghost-gluon vertex. See also below.}. Finally, $\alpha_\Delta$ will be set to zero, to incorporate the corresponding non-renormalization theorem, which at least exists for the limiting interpolating gauge to Coulomb gauge \cite{Fischer2005}. Of course, the latter is an assumption at this stage. The desired solution should have the property $\kappa_V=1$ in four dimensions ($s=3$) and $\kappa_V=1/2$ in three dimensions ($s=2$) to obtain an appropriate potential.

With power-law ans\"atze and the dimensional renormalization, the integrals can be evaluated analytically. The ghost equation \pref{dsghosthor} then yields
\bea
&&\frac{1}{b_\Delta^2 b_T a_\alpha}p^{2+2\kappa_\Delta}=p^{s-2-2\kappa_T+2\alpha_T-2\kappa_\Delta}\times\\
&&\times\frac{g^2N(s-1)}{2^{s+1}\pi^s}\frac{\Gamma(\frac{s}{2}-1-\kappa_T+\alpha_T)\Gamma(\frac{s}{2}-\kappa_\Delta)\Gamma(2-\frac{s}{2}+\kappa_T-\alpha_T+\kappa_\Delta)}{\Gamma(2+\kappa_T-\alpha_T)\Gamma(s-1-\kappa_T+\alpha_T-\kappa_\Delta)\Gamma(1+\kappa_\Delta)},\nn
\eea
\no where again the horizon condition has been implemented. The equation \pref{dsdazeroazeroh}, leads to 
\bea
&&\frac{1}{b_\Delta^2b_Ta_\alpha}p^{2-2\kappa_V+4\kappa_\Delta}=p^{2}+p^{s-2+2\alpha_T-2\kappa_T-2\kappa_V}\times\\
&&\times\frac{g^2 N(s-1)}{2^{s+1}\pi^s}\frac{\Gamma(-1+\frac{s}{2}+\alpha_T-\kappa_T)\Gamma(\frac{s}{2}-\kappa_V)\Gamma(2-\frac{s}{2}-\alpha_T+\kappa_T+\kappa_V)}{\Gamma(2-\alpha_T+\kappa_T)\Gamma(s-1+\alpha_T-\kappa_T-\kappa_V)\Gamma(1+\kappa_V)}\nn.
\eea
\no Finally, the equation \pref{cancelir} yields the lengthy expression \bea
&&4b_T^2p^{-2-4\kappa_T}-1=\\
&&-\frac{g^2N a_\beta b_T b_0}{16\pi^{-2s}}\frac{\Gamma(\frac{s}{2}-1+\beta_0-\kappa_V)}{\Gamma(1-\beta_0+\kappa_V)}p^{-4+s+2\beta_0+2\beta_\pi+2\beta_T-\kappa_T-\kappa_V}\nn\\
&&\Big((2^{2-s}\pi^s\Gamma(\frac{s}{2}-1+\beta_T-\kappa_T)\Gamma(2-\frac{s}{2}-\beta_0-\beta_T+\kappa_T+\kappa_V)\nn\\
&&\times(6-3s+4\beta_T^2+\beta_0(4\beta_T-2-4\kappa_T)+4\beta_T(-3+s-2\kappa_t-\kappa_V)+2\kappa_V\nn\\
&&+4\kappa_T(3-s+\kappa_T+\kappa_V)))\nn\\
&&\times\frac{1}{\Gamma(2-\beta_T+\kappa_T)\Gamma(s-1+\beta_0+\beta_T-\kappa_T-\kappa_V)}\nn\\
&&+\frac{\Gamma(\frac{s}{2}+\beta_\pi+\kappa_T)\Gamma(-1-\frac{s}{2}-\beta_0-\beta_T-\kappa_T+\kappa_V)}{4^s\Gamma(-\beta_\pi-\kappa_T)\Gamma(1+s+\beta_0+\beta_\pi+\kappa_T-\kappa_V)}\nn\\
&&\Big(2^{2+s}\pi^d\beta_0^2(3+2\beta_\pi+2\kappa_T)(s+2\beta_\pi+2\kappa_T)+(2\pi)^s(s+2\beta_\pi+2\kappa_T)\nn\\
&&\times(-2+5s+4\beta_\pi^2+4\kappa_T(1+s+\kappa_T-\kappa_V)+4\beta_\pi(1+s+2\kappa_T-\kappa_V)-6\kappa_V)\nn\\
&&\times(2+s+2\beta_\pi+2\kappa_T-2\kappa_V)+4\beta_0(2^{1+s}s(1+2s)\pi^s+3s(5+s)(2\pi)^s(\beta_\pi+\kappa_T)\nn\\
&&+2^{1+s}\pi^s((\beta_\pi+\kappa_T)(2+4\beta_\pi^2+\kappa_T(7+5d+4\kappa_T)+\beta_\pi(7+5s+8\kappa_T))\nn\\
&&-(3+2\beta_\pi+2\kappa_T)(s+2\beta_\pi+2\kappa_T)\kappa_V))\Big)\Big)\Big).
\eea
\no The characteristics of these equations are better visible in the form
\bea
\frac{1}{b_\Delta^2 b_T a_\alpha}p^{2+2\kappa_\Delta}&=&f(s,\alpha_T-\kappa_T,\kappa_\Delta)p^{s-2-2\kappa_T+2\alpha_T-2\kappa_\Delta}\label{gce}\\
\frac{1}{b_\Delta^2b_Ta_\alpha}p^{2-2\kappa_V+4\kappa_\Delta}&=&p^{2}+f(s,\alpha_T-\kappa_T,\kappa_V)p^{s-2+2\alpha_T-2\kappa_T-2\kappa_V}\label{pce}\\
4b_T^2p^{-2-4\kappa_T}&=&1+h(s,\beta_T,\beta_0,\beta_\pi,\kappa_V,\kappa_T)p^{-4+s+2\beta_0+2\beta_\pi+2\beta_T-\kappa_T-\kappa_V}\\
\label{tce}
\eea
\no The appearance of the same function $f$ in both equations \pref{gce} and \pref{pce} is not a coincidence but due to the same type of coupling and integral kernel appearing. This is entirely due to the truncation, which implied tree-level tensor structures and the same dressing function for both. Of course, an immediate solution would be then $\kappa_V=\kappa_\Delta$, but unfortunately, this is forbidden by integral convergence \pref{boundk}.

Furthermore, the appearance of the dressing of the ghost and $\phi$-vertices only shift the gluon exponent in the respective equations by $\alpha_T$, but yields no modification of neither the exponent nor the pre-factor consistency conditions. Only in the $A_T$ equation \pref{tce} a consequence would be possible, but it will turn out that this is of minor relevance.

\begin{figure}
\epsfig{file=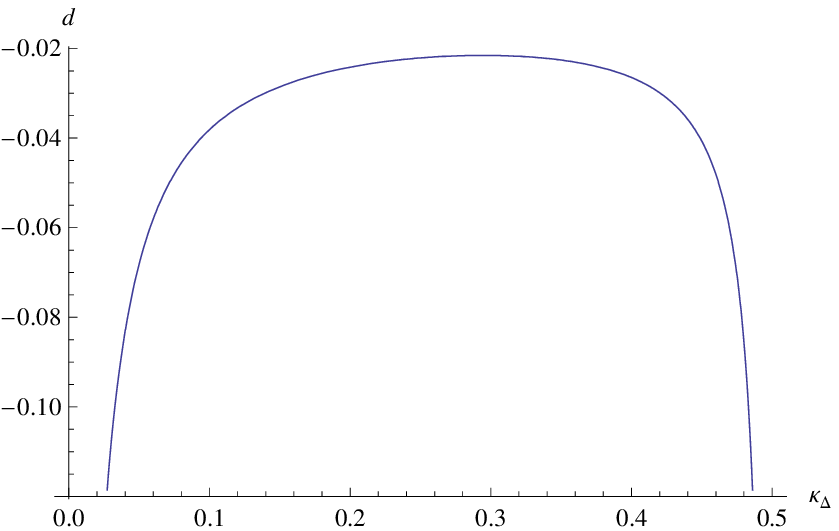,width=\linewidth}\\
\epsfig{file=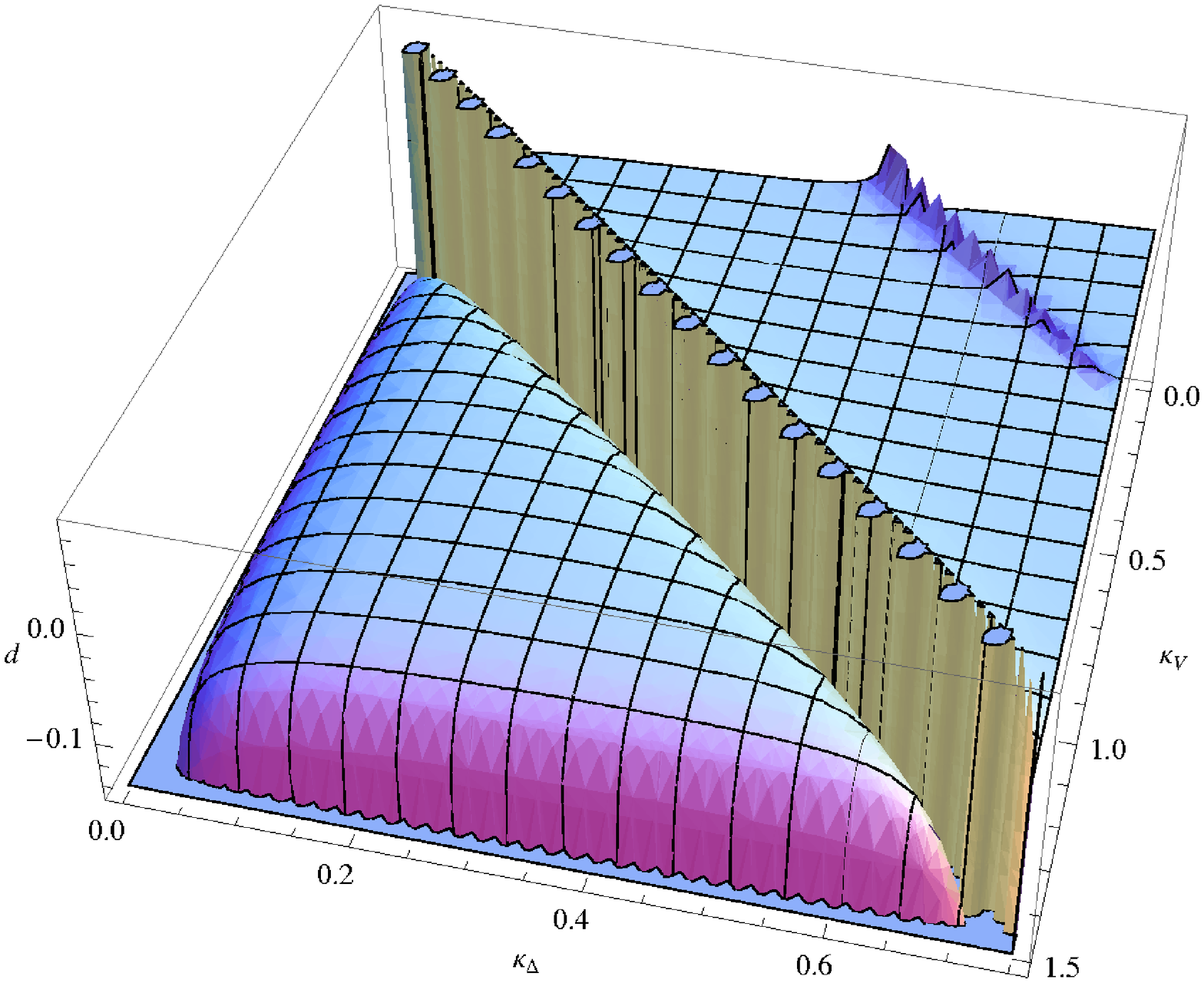,width=\linewidth}
\caption{\label{std} Top panel: The function $d=f(3,-1/2-2\kappa_\Delta,\kappa_\Delta)-f(3,-1/2-2\kappa_\Delta,1)$ within the permitted range $0<\kappa_\Delta<1/2$ for $\kappa_V=1$. A zero crossing would indicate a solution to the consistency condition. Bottom panel: The function $d=f(3,-1/2-2\kappa_\Delta,\kappa_\Delta)-f(3,-1/2-2\kappa_\Delta,\kappa_V)$. The permitted ranges for $\kappa_\Delta$ and $\kappa_V$ is the lower left triangle of the plot, i.\ e., the part including the origin up to the wall-like structure.}
\end{figure}

Let us start with the case $s=3$. First of all, the consistency of the exponents on both sides of \pref{pce} yields the same relation as from the equation \pref{gce},
\be
0=\frac{s-4}{2}-\kappa_T+\alpha_T-2\kappa_\Delta\nn,
\ee
\no as $\kappa_V$ drops out of both sides of \pref{pce}. For the tree-level term in equation \pref{pce} to be subleading, it is required that $2\kappa_\Delta<\kappa_V$, if both are positive. Trying to solve for the pre-factor consistency condition with this limitation, and the limitations $0<\kappa_\Delta<\min(1,\kappa_V/2)$ from integral convergence of \pref{gce} and \pref{pce} and $\kappa_V<s/2$ from equation \pref{pce}, yield that again no solution in three spatial dimensions exists. This is directly visible from the figure \ref{std}, left panel. Only if one allows the ghost and the $A_0\phi$ vertices to vary independently, is it possible to find a solution. This is then also consistent with the equation \pref{tce}. However, it is no longer self-consistent. So this is the unfortunate case, where a solution by power-counting exists, but there is however no solution of the system.

From figure \ref{std}, right panel,  it is clear that relaxing the restriction $\kappa_V=1$ is not sufficient to find a solution: In all cases permitted by integral convergence, no solution exists. Only in the sense of section \ref{scrit}, if one would permit a deviation of $s$ from 3, a solution could be found in the same manner.

In principle, it is possible to relax the integral convergence condition for the equation \pref{pce}, and admit a logarithmic divergence. In this case, $\kappa_V=2\kappa_\Delta$ is also possible. This automatically yields that by power-counting all terms scale as the tree-level term. It can therefore no longer be dropped. However, a possible renormalization constant modifies the value of the tree-level constant also by finite parts. This in turn modifies the consistency condition even further. It is then possible to find a solution for the equations \pref{gce} and \pref{pce}. Selecting, e.\ g., that the finite part of the tree-level term is not modified, and thus a MS-like renormalization scheme, yields a solution as $(\kappa_\Delta,\kappa_V)\approx (0.66645,1.3329)$.

However, even then the case $\kappa_V=1$ remains pathological, because the function $f$ diverges for the required value $\kappa_\Delta=1/2$.

In addition, the case $\kappa_V=2\kappa_\Delta$ would be due to IR-UV mixing, as the value of the exponent becomes entirely determined by the renormalization prescription, which seems undesirable. Still, it would be a solution, and the remaining equation, \pref{tce}, can then be solved with the freedom of four additional parameters, $\beta_T$, $\beta_0$, $\beta_\pi$, and $\kappa_T$, since only the combination $\alpha_T-\kappa_T$ is restricted by \pref{gce} and \pref{pce}.

Hence, there seems to be no pure infrared solution to the system in this truncation in a self-consistent manner at $s=3$. Introduction of additional vertex tensor structures or relaxing some assumptions may be sufficient, but this is currently unclear.

Finally, note that if one accepted a non-self-consistency in the vertex ans\"atze, there would be solutions with $\kappa_V=1$ for appropriate different choices of the exponents $\alpha_T$ in \pref{gce} and \pref{pce} and of the other exponents in \pref{tce}. In this case, even a bare ghost-gluon vertex would be permitted.

The situation in three dimensions, $s=2$, is again different. There is still no solution for $\kappa_V\le 1/2$. However, for $\kappa_V>1/2$, there is a solution, with $\kappa_\Delta$ starting to grow from zero at $\kappa_V=1/2$. By adjusting $\alpha_T$ appropriately, it is then possible to have a vanishing gluon propagator even at $\kappa_V$ only marginally greater than 1/2. In addition, the equation \pref{tce} is then easily solved, if $h$ is positive for some set of the parameters, which can be be achieved even at $\beta_T=\beta_\pi=\beta_0=0$. However, a massive gluon is then impossible, as for $\kappa_T=-1$ the function $h$ vanishes.

So, neither in four nor in three space-time dimensions in this truncation scheme does the desired solution exist. In three space dimensions no solution exists at all, and in two space dimensions only a solution with $D_{00}$ that is too divergent. 

\section{Summary}\label{ssum}

In the present work we have studied the infrared limit of the DSEs in the first-order formalism of Coulomb-gauge Yang-Mills theory. Our truncation was based on the assumption of dominance by the instantaneous contributions. However, it turned out that, even taking into account a certain class of non-trivial vertices, no solution was found which directly exhibited the desired linear rising Coulomb string tension. However, once this condition is relaxed, solutions can be found. At non-vanishing temperatures it is quite reasonable to alleviate some of the conditions, corresponding approaches are discussed in \cite{kldz}.

We also think that the BRST-type on-shell operator $r$ introduced in section \ref{spsa} may be useful in future studies of the Coulomb gauge.

The results presented here will be useful in constructing other truncation schemes, which may be better able to capture the infrared dynamics in this formalism. From our discussions it has become clear that such a truncation scheme has to go beyond the instantaneous approximation. This then also constitutes the main result of our investigation.

\smallskip
{\bf Acknowledgments}\\

This work is dedicated to Prof.\ Willibald Plessas on the occasion of his 60th birthday.

We are grateful to Peter Watson for a critical reading of the manuscript and helpful comments.

This work was supported by the FWF under grant numbers P20330 and M1099-N16 and by the DFG under grant number Ma 3935/1-1 and Ma 3935/1-2 (AM).

\end{document}